\documentclass{article}
\usepackage{spconf,amsmath,graphicx}

\DeclareMathOperator*{\argmax}{arg\,max}
\usepackage{algorithm}
\usepackage[noend]{algpseudocode}
\usepackage{mathtools}
\usepackage{url}
\usepackage{makecell}
\usepackage{subcaption}
\usepackage{amssymb}
\usepackage{amsfonts}
\usepackage{dsfont}
\usepackage{amsthm}
\usepackage[super]{nth}
\usepackage{lipsum}
\usepackage{etoolbox}

\theoremstyle{definition}

\title{Detecting Backdoor Attacks against Point Cloud Classifiers\vspace{-0.125in}}
\name{\vspace{-0.16in}Zhen Xiang$^1$, David J. Miller$^1$, Siheng Chen$^2$, Xi Li$^1$ and George Kesidis$^1$
\address{$^1$Pennsylvania State University, $^2$Shanghai Jiao Tong University}
}

\begin{document}
%
\maketitle
\begin{abstract}
\vspace{-0.05in}
Backdoor attacks (BA) are an emerging threat to deep neural network classifiers. A classifier being attacked will predict to the attacker's target class when a test sample from a source class is embedded with the backdoor pattern (BP). Recently, the first BA against point cloud (PC) classifiers was proposed, creating new threats to many important applications including autonomous driving. Such PC BAs are not detectable by existing BA defenses due to their special BP embedding mechanism. In this paper, we propose a reverse-engineering defense that infers whether a PC classifier is backdoor attacked, without access to its training set or to any clean classifiers for reference. The effectiveness of our defense is demonstrated on the benchmark ModeNet40 dataset for PCs.
\end{abstract}
\vspace{-0.05in}
\vspace{-0.1in}
\section{Introduction}
\label{sec:intro}
\vspace{-0.1in}
Deep neural network classifiers have achieved good performance in many point cloud (PC) classification tasks \cite{3D_review, Robotics}. However, they are shown to be vulnerable to adversarial attacks \cite{Review, 3D_TTE_Seminal, HighSchoolKid}, including a recently proposed PC backdoor attack (BA) \cite{PCBA}. BAs were initially proposed for the image domain. A classifier being attacked will likely predict to the attacker's {\it target class} whenever a test sample from a {\it source class} of the attack is embedded with a {\it backdoor pattern} (BP) \cite{Targeted, Trojan, BAsurvey}. BAs are also not easily detectable since they negligibly affect the classifier's accuracy on clean test samples.

Defenses against BAs have been extensively studied for images. Existing state-of-the-art BA defenses mostly belong to a category of {\it reverse-engineering defenses} (REDs). REDs detect whether a classifier is backdoor attacked without access to its training set and without reference to any clean classifiers trained for the same domain \cite{NC, Tabor, Post-TNNLS, RED_black_box, DeepInspect, MAMF}. These advantages make REDs suitable for the most popular, practical scenario nowadays, where training of the classifier is outsourced due to high computational cost and the need for ``big data'' for training \cite{BadNet}. In such a scenario, the defender is merely the consumer of the classifier (e.g. a mobile app user) without access to the training process and with no capability to train clean classifiers for the same domain \cite{DataLimited}.

Despite the success of existing REDs against image BAs, they are not applicable to the recently proposed PC BA. To facilitate practical implementation using physical objects (e.g. a ball carried by a pedestrian), the BP for this PC BA is designed as a small set of points inserted at a {\it common} spatial location {\it close} to the original points of all source class PCs \cite{PCBA}. However, typical BPs for image BAs are either tiny additive perturbations \cite{Post-TNNLS} or small patch triggers \cite{NC}. This difference in BP type prevents existing BP reverse-engineering formulated for image REDs from being applied to PCs.

In this paper, we propose a RED to detect whether a PC classifier is backdoor attacked -- this is the {\it first} defense against PC BAs {\it without} access to the classifier's training set. We propose a novel BP reverse-engineering problem specific to PCs. Different from image REDs, which trial-estimate a BP either for each putative target class \cite{NC, L-RED} or for each putative (source, target) class pair \cite{Post-TNNLS, MAMF} we perform BP reverse-engineering for each putative {\it source class} and simultaneously estimate a target class. This different design (compared with image REDs) addresses the generally strong robustness of PC classifiers \cite{PointNet, 3D_TTE_Seminal}, for which BP estimation is a hard problem for a non-negligible number of putative target classes. Moreover, for some putative source classes, there exists a spatial location {\it close to} most source class PCs, such that a single inserted point can cause most of these PCs to be misclassified to a {\it common} target class, {\it irrespective of the existence of a BA}. Such an ``{\it intrinsic backdoor}'' can easily cause false detections when there is actually no attack. While distinguishing intrinsic backdoors from those caused by attack is still an open problem even for image BA defense \cite{Post-TNNLS}, we propose a novel combined detection statistic to address this challenging problem. Finally, we evaluate our detector on the benchmark ModeNet40 dataset for PCs to show its effectiveness.

\vspace{-0.1in}
\section{Related Work}\label{sec:related_work}
\vspace{-0.1in}
\subsection{BA against PC Classifiers}\label{subsec:PCBA}
\vspace{-0.05in}
Consider a PC domain with sample space ${\mathcal X}$ and label space ${\mathcal C}$. A PC BA aims to have the victim classifier $f(\cdot):{\mathcal X}\rightarrow{\mathcal C}$ predict to some attacker's target class $t^{\ast}\in{\mathcal C}$ whenever a test sample ${\bf X}=\{{\bf x}_i\in{\mathbb R}^3|i=1, \cdots, n\}\in{\mathcal X}$ from some source class $s^{\ast}\in{\mathcal C}$ is embedded with a BP ${\bf V}^{\ast}$ \cite{PCBA}. Here, ${\bf V}^{\ast}$ is a set of {\it inserted points} jointly specified by a spatial location ${\bf c}^{\ast}\in{\mathbb R}^3$ and a local geometry ${\bf U}^{\ast}=\{{\bf u}^{\ast}_j\in{\mathbb R}^3 | j=1, \cdots, n'\}$:
\begin{equation}\label{eq:BP_def}
\vspace{-0.05in}
{\bf V}^{\ast} = \{{\bf u}^{\ast}_j + {\bf c}^{\ast}|{\bf u}^{\ast}_j\in{\mathbb R}^3, {\bf c}^{\ast}\in{\mathbb R}^3, j=1, \cdots, n'\}.
\end{equation}
For ${\bf V}^{\ast}$, the spatial location ${\bf c}^{\ast}$ is optimized by the attacker such that its distance to these class $s^{\ast}$ PCs, measured by ${\mathbb E}_{{\bf X}\sim P_{s^{\ast}}} [d({\bf c}^{\ast}, {\bf X})]$, is sufficiently small, where $d({\bf c}, {\bf X}) = \min_{{\bf x}\in{\bf X}}||{\bf c}-{\bf x}||_2$
is the distance between ${\bf c}\in{\mathbb R}^3$ and ${\bf X}\in{\mathcal X}$. $P_k$ is the sample distribution for class $k\in{\mathcal C}$. Then, a {\it perfectly} successful {\it backdoor mapping} would achieve $f({\bf X}\cup{\bf V}^{\ast})=t^{\ast}$, $\forall{\bf X}\sim P_{s^{\ast}}$;
and with no misclassifications induced for samples not from a source class of the attack. Like image BAs, a PC BA is typically launched by poisoning the classifier's training set with a small set of PCs originally from class $s^{\ast}$, embedded with the same BP ${\bf V}^{\ast}$, and labeled to class $t^{\ast}$. PC BAs are also hard to detect since they have negligible effect on classifier's predictions for PCs with BP.

\vspace{-0.1in}
\subsection{BA Defense}\label{subsec:BA_defense}

BA defenses have been extensively studied for images; but no defenses have been proposed for PC BAs yet. Some BA defenses aim to find and remove poisoned samples (with BP embedded) from the training set \cite{SS, AC, CI}. Irrespective of their deployment being infeasible for scenarios where the defender is the user of the classifier without access to the training set, these defenses are anyway not effective for PC BAs \cite{PCBA}.

A family of {\it reverse-engineering defenses} does not require access to the classifier's training set. They trial-estimate a BP for each putative (source, target) class pair (or target class only \cite{NC, L-RED}) using a small, clean dataset independently collected by the defender \cite{Post-TNNLS}. When there is an attack, the pattern estimated for the true BA class pair should be related to the BP used by the attacker and exhibit some atypicality compared with patterns estimated for non-BA class pairs. For example, to detect image BAs with an additive perturbation BP, \cite{Post-TNNLS} builds a purely unsupervised anomaly detector based on the fact that the norm of such BP is much smaller than the minimum perturbation norm required to induce high group misclassification for non-BA class pairs. Thus, a BA is detected when there exists a class pair with an abnormally small estimated perturbation norm. However, REDs are tailored to the BP embedding mechanism; thus, existing REDs designed for images BAs are not applicable to PC BAs.

\vspace{-0.1in}
\section{Method}\label{sec:method}
\vspace{-0.1in}
\subsection{Overview}\label{subsec:overview}
\begin{figure}[t]
	\vspace{-0.05in}
	\centering
	\begin{minipage}[b]{.31\linewidth}
		\centering
		\centerline{\includegraphics[width=\linewidth]{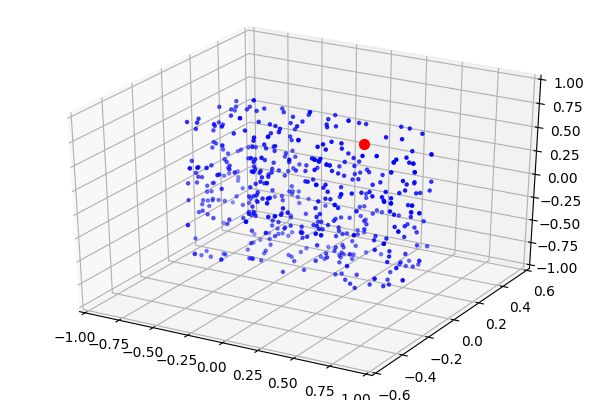}}
		\vspace{-0.05in}
	\end{minipage}
	\begin{minipage}[b]{.31\linewidth}
		\centering
		\centerline{\includegraphics[width=\linewidth]{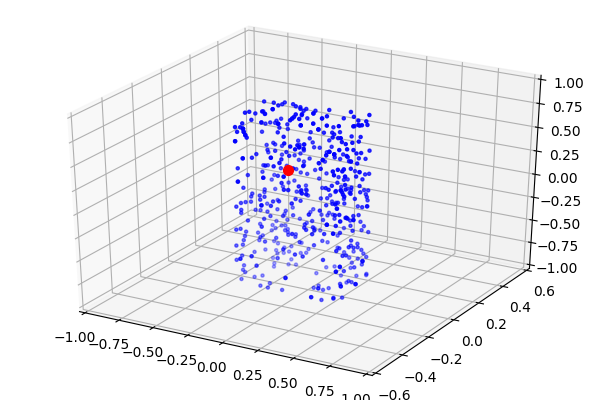}}
		\vspace{-0.05in}
	\end{minipage}
	\begin{minipage}[b]{.31\linewidth}
		\centering
		\centerline{\includegraphics[width=\linewidth]{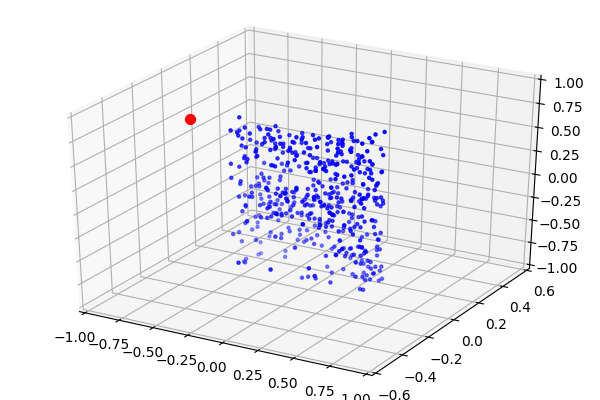}}
		\vspace{-0.05in}
	\end{minipage}
	\caption{Example of intrinsic backdoor from one of our experiments (${\rm P}_6$-PN in Tab \ref{tab:order_stat_pvs}): for PCs from the same source class, spatial locations estimated {\it sample-wise} (in red) are all close to these PCs (in blue), but are different from PC to PC.}
	\label{fig:intrinsic_backdoor}
\end{figure}

\vspace{-0.05in}
{\bf Goals and assumptions.} The defender aims to infer whether a classifier is backdoor attacked and to determine the target class if an attack is detected. The defender has no access to the classifier's training set and is not capable of training any classifiers. The defender does possess an independently collected small, clean dataset for detection. These goals and assumptions are the same as for existing image REDs \cite{Post-TNNLS, NC, Tabor}.\\
{\bf Key ideas.} Our detector is based upon the following intuitions. These intuitions not only guide our detector design, but are also verified experimentally by the success of our detector. {\bf I1}: For {\it most} non-BA class pairs, a {\it common} set of inserted points that induces high group misclassification from source class to target class will be spatially {\it far} from the points of source class PCs; but for BA class pair $(s^{\ast}, t^{\ast})$, the existence of the {\it backdoor mapping} guarantees the existence of a {\it common} spatial location close to the source class PCs (likely near ${\bf c}^{\ast}$), where a set of inserted points can induce most source class PCs to be misclassified to the target class. {\bf I2}: A few non-BA class pairs may be associated with an {\it ``intrinsic backdoor''}. For these class pairs, like a true BA class pair, there exists a spatial location close to most PCs from the source class, such that a common set of inserted points there will induce most of these PCs to be misclassified to the target class. However, such a spatial location, different from ${\bf c}^{\ast}$ (specified by the attacker) for the true backdoor, will likely be {\it close to} the points of most {\it target class} PCs. {\bf I3}: Unlike backdoor mappings caused by attack with a {\it single common} spatial location ${\bf c}^{\ast}$, an intrinsic backdoor is likely due to the source and target classes being {\it ``semantically''} similar, such that there may exist {\it several} intrinsic backdoor points for a given non-BA class pair, with each one close to source class PCs (Fig. \ref{fig:intrinsic_backdoor}). In this case (for a source class with an intrinsic backdoor), the {\it closest} ({\it sample-wise}) spatial location for a set of inserted points to induce a ({\it sample-wise}) misclassification to the target class can be different for different PCs from the same source class.\\
{\bf Detection procedure.} Our detection method consists of a BP estimation step followed by a detection inference step.

\vspace{-0.1in}
\subsection{Step 1: BP Estimation}\label{subsec:BP_estimation}

To find BA class pairs if there are any, based on I1, we need to first find, for each class pair, the {\it common} spatial location {\it closest} to the source class PCs such that a set of points inserted there induces most of these PCs to be misclassified to the target class, i.e., we need to reverse-engineer the BP. According to \cite{PCBA}, the backdoor mapping mostly relies on the spatial location ${\bf c}^{\ast}$ but not the local geometry ${\bf U}^{\ast}$ of the inserted points. Thus, we can focus on reverse-engineering the spatial location of BP with arbitrary local geometry -- for simplicity, we insert a {\it single} point at the spatial location. Formally, we aim to solve the following problem for each class pair $(s, t)\in{\mathcal C}\times{\mathcal C}$:
\begin{equation}\label{eq:opt_raw}
\vspace{-0.15in}
\begin{aligned}
& \underset{{\bf c}\in{\mathbb R}^3}{\text{min}}
& & \sum_{{\bf X}\in{\mathcal D}_s} d({\bf c}, {\bf X})\\
& \text{s. t.}
& & \frac{1}{|\mathcal{D}_s|}\sum_{{\bf X}\in{\mathcal D}_s} \mathds{1}\big[ f({\bf X}\cup\{{\bf c}\})=t \big]\geq \pi,
\end{aligned}
\end{equation}
where ${\mathcal D}_s$ is the subset of clean samples from class $s$ possessed by the defender; ${\mathds 1}[\cdot]$ is the indicator function; and $\pi\in[0, 1]$ is a group misclassification fraction which is typically set large (e.g. $\pi=0.9$ was chosen in \cite{Post-TNNLS}).

However, problem (\ref{eq:opt_raw}) is difficult to solve in practice. First, the indicator function in the constraint is not differentiable. Second, unlike image BPs (e.g. an additive perturbation) typically constrained by some range of valid pixel values, the search space for a point spacial location is unlimited. Also, due to the generally strong adversarial robustness of recent PC classifiers \cite{PointNet, 3D_TTE_Seminal}, for many class pairs, there may not even exist a spatial location reasonably close to the source class PCs, where an inserted point can induce high (e.g. at least $\pi$) group misclassification to the target class. For these class pairs, even finding a solution just to satisfy the constraint of (\ref{eq:opt_raw}) may be infeasible in practice. To address the two challenges above, we perform BP estimation for each putative {\it source class} by minimizing a {\it differentiable surrogate} objective. In particular, for each source class $s\in{\mathcal C}$, we search for the closest spatial location to PCs from class $s$, such that a point inserted there causes at lease $\pi$ fraction of these PCs to be misclassified to any class {\it other than} $s$ (untargeted misclassifications). Formally, we minimize loss:
\begin{equation}\label{eq:loss}
\vspace{-0.1in}
\begin{aligned}
\hspace{-0.05in} l({\bf c}; {\mathcal D}_s, \lambda) \hspace{-0.025in} = \hspace{-0.025in} &\sum_{{\bf X}\in{\mathcal D}_s} \big[ h(s|{\bf X}\cup\{{\bf c}\}) - \underset{k\neq s}{\text{max}} \, h(k|{\bf X}\cup\{{\bf c}\}) \big]\hspace{-0.05in} \\
&+ \lambda \sum_{{\bf X}\in{\mathcal D}_s} d({\bf c}, {\bf X})
\end{aligned}
\end{equation}
over ${\bf c}$ using Alg. \ref{alg:BP_estimation}. The first sum in Eq. (\ref{eq:loss}) is inspired by the {\it untargeted} CW loss in \cite{CW}, where $h(k|{\bf X})$ is the output of ${\bf X}\in{\mathcal X}$ for class $k\in{\mathcal C}$ directly prior to the softmax activation, which is smoother for optimization than the class posterior constrained in interval $[0, 1]$. Using such untargeted loss, we let the source class PCs ``vote'' for a target class by:
\begin{equation}\label{eq:target_class}
\vspace{-0.1in}
\hat{t}(s) = \underset{k\neq s}{\text{argmax}} \, \sum_{{\bf X}\in{\mathcal D}_s} {\mathds 1}\big[ f({\bf X}\cup\{\hat{\bf c}(s)\})=k \big]
\end{equation}
where $\hat{\bf c}(s)$ is the spatial location estimated for class $s$. Also, compared with some image REDs that estimate BPs for each class pair \cite{Post-TNNLS}, our detector performs ${\mathcal O}(K)$ BP estimations instead of ${\mathcal O}(K^2)$ (where $K=|{\mathcal C}|$ is the number of classes); thus it is more {\it efficient} for large $K$. The second sum in Eq. (\ref{eq:loss}) is a regularizer which constrains the distance of ${\bf c}$ to the points of the source class PCs. The coefficient $\lambda$ is automatically adjusted according to line 6-10 of Alg. \ref{alg:BP_estimation}.

\begin{algorithm}[t]
	\footnotesize
	\caption{Spatial location estimation for class $s\in{\mathcal C}$.}\label{alg:BP_estimation}
	\begin{algorithmic}[1]
		\State {\bf Inputs}: data subset ${\mathcal D}_s$, classifier $f(\cdot)$, fraction $\pi$, step size $\delta$, maximum iteration count $\tau_{\rm max}$, scaling factor $\alpha$.
		\State {\bf Initialization}: ${\bf c}^{(0)}\sim{\mathcal N}({\mathbf 0}, {\bf I})$, $\lambda^{(0)}$ set to a small positive number (e.g. $10^{-5}$), $\hat{\bf c}(s)={\boldsymbol{\infty}}$, $\rho^{(0)}=0$.
		\For{$\tau = 0:\tau_{\rm max}-1$}
		\State ${\bf c}^{(\tau+1)} = {\bf c}^{(\tau)} - \delta \nabla_{\bf c} l({\bf c}; {\mathcal D}_s, \lambda^{(\tau)}) |_{{\bf c}={\bf c}^{(\tau)}}$
		\State $\rho^{(\tau+1)}=\frac{1}{|\mathcal{D}_s|}\sum_{{\bf X}\in{\mathcal D}_s} {\mathds 1}\big[ f({\bf X}\cup\{{\bf c}^{(\tau+1)}\}) \neq s \big]$
		\If{$\rho^{(\tau+1)}\geq\pi$}
		\State $\lambda^{(\tau+1)} = \lambda^{(\tau)}\cdot\alpha$
		\If{$\sum_{{\bf X}\in{\mathcal D}_s} \big[d({\bf c}^{(\tau+1)}, {\bf X}) - d(\hat{\bf c}(s), {\bf X})\big]<0$}
		\State $\hat{\bf c}(s) = {\bf c}^{(\tau+1)}$
		\EndIf
		\State {\bf else} $\lambda^{(\tau+1)} = \lambda^{(\tau)}/\alpha$
		\EndIf
		\EndFor
		\State {\bf Outputs}: $\hat{\bf c}(s)$
	\end{algorithmic}
\end{algorithm}

In addition to the {\it group} BP estimation above, based on intuition I3, for each putative source class $s\in{\mathcal C}$, we also need to estimate a {\it sample-wise} spatial location for each ${\bf X}\in{\mathcal D}_s$, given the estimated target class $\hat{t}(s)$. Formally, we minimize the following loss using the same Alg. \ref{alg:BP_estimation}:
\begin{equation}\label{eq:loss_sample}
\tilde{l}({\bf c}; {\bf X}, \lambda) = h(s|{\bf X}\cup\{{\bf c}\}) - h(\hat{t}(s)|{\bf X}\cup\{{\bf c}\})
+ \lambda d({\bf c}, {\bf X})
\end{equation}
with ${\mathcal D}_s$ replaced by a set of a single PC $\{{\bf X}\}$, and with the loss in line 4 replaced by Eq. (\ref{eq:loss_sample}). We denote the estimated sample-wise (SW) spatial location for ${\bf X}\in{\mathcal D}_s$ as $\hat{\bf c}_{\rm sw}(s, {\bf X})$.

\begin{table*}
	\vspace{-0.45in}
	\begin{center}
		\renewcommand{\arraystretch}{1.1}
		\resizebox{\textwidth}{!}{
			\begin{tabular}{ c|ccccccccc }
				\hline
				& ${\rm P}_1$-PN & ${\rm P}_2$-PN & ${\rm P}_3$-PN & ${\rm P}_4$-PN & ${\rm P}_5$-PN & ${\rm P}_6$-PN & ${\rm P}_7$-PN & ${\rm P}_1$-PN++ & ${\rm P}_1$-DGCNN \\
				\hline
				$1/r_{\rm s}$ & (6.2e\textsuperscript{-3}, 0.36) & (3.8\textsuperscript{-3}, 0.16) & (4.3e\textsuperscript{-15}, 0.33) & (2.2e\textsuperscript{-7}, 2.6e\textsuperscript{-2}) & (0.24, 0.11) & (0.24, 1.6e\textsuperscript{-2}) & (4.3e\textsuperscript{-3}, 9.7e\textsuperscript{-2}) & (u.f., 8.2e\textsuperscript{-6}) & (4.4e\textsuperscript{-5}, 4.3e\textsuperscript{-2})\\
				$r_{\rm t}/r_{\rm s}$ & (4.5e\textsuperscript{-2}, 9.2e\textsuperscript{-6}) & (u.f., 0.32) & (6.1e\textsuperscript{-6}, 9.8e\textsuperscript{-2}) & (2.8e\textsuperscript{-3}, 0.58) & (0.12, 0.19) & (0.21, 0.60) & (6.7e\textsuperscript{-5}, 9.0e\textsuperscript{-3}) & (u.f., 0.99) & (0.10, 0.59)\\
				$w/r_{\rm s}$ & (1,7e\textsuperscript{-7}, 0.19) & (3.5e\textsuperscript{-3}, 0.26) & (u.f., 0.27) & (5.6e\textsuperscript{-9}, 9.2e\textsuperscript{-3}) & (1.4e\textsuperscript{-2}, 6.1e\textsuperscript{-2}) & (1.4e\textsuperscript{-2}, 2.6e\textsuperscript{-2}) & (5.5e\textsuperscript{-9}, 7.0e\textsuperscript{-3}) & (u.f., 0.94) & (0.22, 2.9e\textsuperscript{-2})\\
				\hline
				$r=w\cdot r_{\rm t}/r_{\rm s}$ & (3.3e\textsuperscript{-3}, 0.38) & (u.f., 0.19) & (u.f., 0.20) & (u.f., 0.22) & (5.4e\textsuperscript{-2}, 0.27) & (7.6e\textsuperscript{-4}, 0.33) & (u.f., 9.3e\textsuperscript{-2}) & (5.5e\textsuperscript{-13}, 0.99) & (1.9e\textsuperscript{-3}, 0.18)\\
				\hline
		\end{tabular}}
		\vspace{-0.1in}
		\caption{Order statistic p-value (pv), in form of (attack pv, clean pv), for nine pairs of classifiers being attacked with the associated clean classifier, for the statistic $r$ (Eq. (\ref{eq:statistic})) used by our detector, and for three alternative statistics ($1/r_{\rm s}$, $r_{\rm t}/r_{\rm s}$, and $w/r_{\rm s}$), as part of an ablation study. Attacks are associated with class pairs ${\rm P}_1, ..., {\rm P}_7$ in \cite{PCBA}; classifier architectures include PointNet (PN), PointNet++ (PN++), and DGCNN. ``u.f.'' represents ``underflow'' for positive number less than $10^{-323}$.}
		\label{tab:order_stat_pvs}
	\end{center}
\vspace{-0.3in}
\end{table*}

\subsection{Step2: Detection Inference}\label{subsec:inference}
\vspace{-0.05in}
Our detection statistic is composed of three basic statistics (obtained from BP estimation for each putative source class) corresponding to the three intuitions in Sec. \ref{subsec:overview} respectively. For each $s\in{\mathcal C}$, we get: 1) the average distance from the estimated spatial location to points of source class PCs: $r_{\rm s}(s) = \frac{1}{|{\mathcal D}_s|} \sum_{{\bf X}\in{\mathcal D}_s} d(\hat{\bf c}(s), {\bf X})$;
2) the average distance from the estimated spatial location to points of PCs from the estimated target class $\hat{t}(s)$: $r_{\rm t}(s) = \frac{1}{|{\mathcal D}_{\hat{t}(s)}|} \sum_{{\bf X}\in{\mathcal D}_{\hat{t}(s)}} d(\hat{\bf c}(s), {\bf X})$;
and 3) a normalized similarity score: $w(s) = (z(s) - \underset{k\in{\mathcal C}}{\text{min}} \, z(k)) \big/ (\underset{k\in{\mathcal C}}{\text{max}} \, z(k) - \underset{k\in{\mathcal C}}{\text{min}} \, z(k))$,
where $z(k)=\frac{1}{|{\mathcal D}_k|} \sum_{{\bf X}\in{\mathcal D}_k} \frac{\hat{\bf c}(k) \cdot \hat{\bf c}_{\rm sw}(k, {\bf X})}{|\hat{\bf c}(k)| \, |\hat{\bf c}_{\rm sw}(k, {\bf X})|}$ is the average cosine similarity\footnote{PCs are usually aligned to the origin for classification.} between the estimated sample-wise spatial location for each ${\bf X}\in{\mathcal D}_k$ and the estimated group spatial location for class $k\in{\mathcal C}$. The normalization limits the similarity score in interval $[0, 1]$ for generalization to different domains.

According to intuition I1, $r_{\rm s}(s)$ will likely be large if $(s, \hat{t}(s))$ is a non-BA class pair; otherwise, $r_{\rm s}(s)$ will likely be small. If for some class $s$, $(s, \hat{t}(s))$ is associated with an intrinsic backdoor mapping, such that $r_{\rm s}(s)$ is abnormally small, based on I2 and I3, $r_{\rm t}(s)$ or $w(s)$ (or both) will likely be abnormally small. Thus, for each putative source class $s\in{\mathcal C}$, we compute the combined detection statistic:
\begin{equation}\label{eq:statistic}
\vspace{-0.1in}
r(s) = w(s) \frac{r_{\rm t}(s)}{r_{\rm s}(s)},
\end{equation}
which will be abnormally large only if $(s, \hat{t}(s))$ is a BA pair.

Our inference is based on an {\it unsupervised} anomaly detection. We check among the statistics for all $s\in{\mathcal C}$ if there exists an abnormally large one. Ideally, like the image RED in \cite{Post-TNNLS}, we can fit a null distribution using all statistics excluding the {\it largest one} and evaluate its {\it atypicality} under the null using the maximum {\it order statistic} p-value, which is generally insensitive to the number of statistics used for detection. However, like image BAs, a PC BA may cause {\it collateral damage}: a backdoor mapping with the same BP may be learned for some class pair $(s, t^{\ast})$ with $s\in{\mathcal C}\setminus\{s^{\ast}, t^{\ast}\}$ \cite{Post-TNNLS}. In such case, there may be {\it multiple} abnormally large statistics associated with {\it the same target class} $t^{\ast}$ but different source classes; thus, the null distribution estimated with only the largest statistic being excluded may be biased. To solve this problem, we exclude statistics for all $s\in{\mathcal C}$ such that $\hat{t}(s)=\hat{t}(s_{\rm max})$ when estimating the null, where $s_{\rm max}=\argmax_{k\in{\mathcal C}} r(k)$. Given all statistics in $[0, \infty)$, similar to \cite{Post-TNNLS}, we choose a single-tailed parametric density form (e.g. Gamma distribution in both our experiments and \cite{Post-TNNLS}) for our null distribution, with cdf $G(\cdot)$, such that any abnormally large statistics (likely corresponding to BA class pairs) will likely appear in the tail (e.g. Fig. \ref{fig:hist_stats}). Then the {\it maximum order statistic p-value} is:
\begin{equation}\label{eq:order_stat_pv}
\vspace{-0.1in}
{\rm pv} = 1 - G(r(s_{\rm max}))^{K-J},
\end{equation}
where $J$ is the number of statistics being excluded when estimating the null distribution. A detection threshold $\phi$ is chosen (e.g. the classical $\phi=0.05$), such that a BA is detected with confidence $1-\phi$ if ${\rm pv}<\phi$. When a BA is detected, $\hat{t}(s_{\rm max})$ is inferred as the target class.

\vspace{-0.15in}
\section{Experiments}\label{sec:experiments}
\vspace{-0.1in}
\subsection{Experiment Settings}\label{exp:exp_settings}
{\bf Dataset:} The benchmark ModelNet40 dataset contains 12311 aligned CAD models from 40 object categories \cite{ModelNet40}. 9843 and 2468 PCs are in the training set and test set, respectively.\\
{\bf Attack configurations:} We consider the seven attacks created in \cite{PCBA} for ModelNet40, where the BP for each attack is a set of points inserted at an optimal spatial location and with a random local geometry (i.e. the ``RS'' geometry in \cite{PCBA}). The source and target class pair for these seven attacks are also specified in \cite{PCBA} and named as ``${\rm P}_1, ..., {\rm P}_7$''. For each attack, 15 source class PCs embedded with the BP and labeled to the target class are used for poisoning the training set.\\
{\bf Classifier, training, and attack effectiveness:} We consider the same state-of-the-art PC classifiers as in \cite{PCBA}. In particular, we consider the same PointNet classifiers \cite{PointNet} trained for all seven attacks, the same PointNet++ classifier \cite{PointNetpp} and the same DGCNN classifier \cite{DGCNN} trained for the attack associated with class pair ${\rm P}_1$, with the training configurations detailed in \cite{PCBA}. All the classifiers being attacked exhibit high attack success rate and almost no degradation in clean test accuracy as reported in \cite{PCBA}. To evaluate false detections of our detector, for each classifier being attacked, we also train a classifier with no BA, using the same training configurations.

\vspace{-0.15in}
\subsection{Detection Performance Evaluation and Results}\label{subsec:exp_eval}

\begin{figure}[t]
	\centering
	\begin{minipage}[b]{.46\linewidth}
		\vspace{-0.15in}
		\centering
		\centerline{\includegraphics[width=\linewidth]{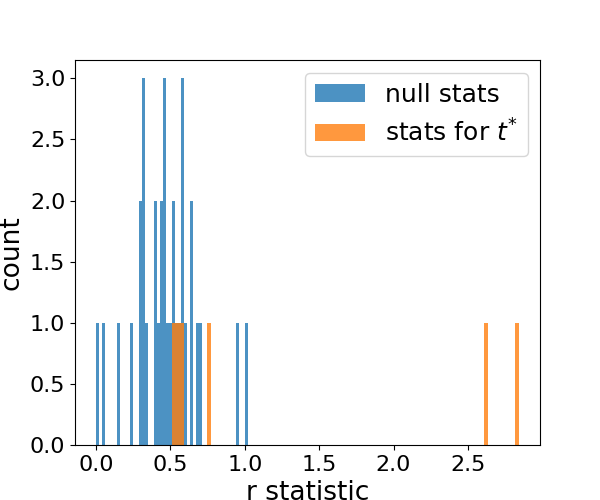}}
		\vspace{-0.1in}
	\end{minipage}
	\begin{minipage}[b]{.46\linewidth}
		\vspace{-0.15in}
		\centering
		\centerline{\includegraphics[width=\linewidth]{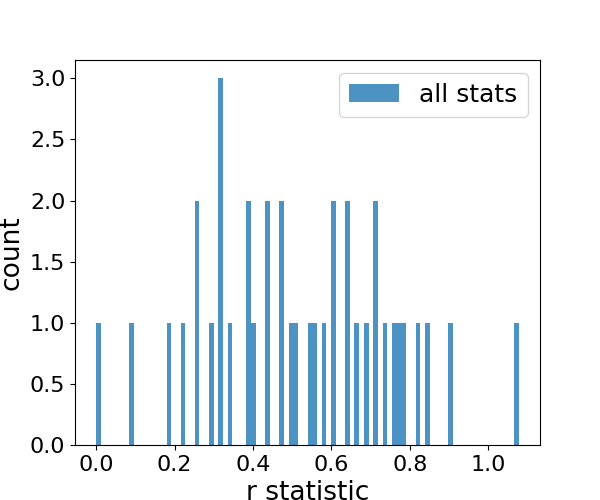}}
		\vspace{-0.1in}
	\end{minipage}
	\caption{Histogram of $r$ statistics for the classifier with BA (left)
	and its associated clean classifier without BA (right). For the attack case, statistics for two source classes ``voting'' to the BA target class $t^{\ast}$ are abnormally large.}
	\label{fig:hist_stats}
\end{figure}

{\bf Detector configurations:} Following the assumptions in Sec. \ref{subsec:overview}, for each class, we randomly select 10 clean PCs that are correctly classified from the ModelNet40 data set, to form the clean set used for detection -- these PCs are not used for training. For BP estimation (both group-wise and sample-wise) using Alg. \ref{alg:BP_estimation}, we set $\pi=0.9$, $\delta=0.1$, $\tau_{\rm max}=3{\rm k}$, and $\alpha=1.5$. These choices are not critical to the performance of our detector, and can be easily chosen to minimize the loss value at convergence. We also perform 10 random initializations and pick the best solution, which is a common practice for solving highly non-convex problems.\\
{\bf Detection performance:} For each classifier being attacked and its associated clean classifier, we report the order statistic p-values obtained using our statistic $r=w\cdot r_{\rm t}/r_{\rm s}$ (Eq. (\ref{eq:statistic})) in Tab. \ref{tab:order_stat_pvs} (last row). In general, the p-values are large for most clean classifiers and are small when there is a BA, as we expect. Applying the classical $\phi=0.05$ threshold to these p-values, {\it we only missed one attack} (${\rm P}_5$-PN, and barely, since the p-value is 0.054) {\it with zero false detections}. For each detected attack, the BA target class is also correctly inferred.\\
{\bf Ablation study:} As the {\it first} defense designed for PC BAs, we do not have a competitor to compare with. Still, we show detection performance using simplified statistics instead of the combined one used by our detector. In Tab. \ref{tab:order_stat_pvs}, for these simplified statistics, the p-values for some clean classifiers (e.g. ${\rm P}_4$-PN, ${\rm P}_6$-PN) are small due to the existence of {\it intrinsic backdoors}, which easily causes false detections. Even for some classifiers being attacked (e.g. ${\rm P}_5$-PN, ${\rm P}_6$-PN), using these simplified statistics, the null estimation will be affected by intrinsic backdoors, such that the resulting p-value will not be small enough to trigger a detection. These results highlight the importance of all three components (each strongly motivated by the intuition in Sec. \ref{subsec:overview}) of our detection statistic.\\
{\bf Visualization of detection:} In Fig. \ref{fig:hist_stats}, we show the distribution of our statistic for both the classifier being attacked and the clean classifier for ${\rm P}_4$-PN, for example. When there is a BA, two statistics associated with the true BA target class clearly appear in the ``tail'' of the null distribution, triggering a detection with correct inference of the BA target class.

\vspace{-0.175in}
\section{Conclusions}\label{sec:conclusions}
\vspace{-0.125in}
We proposed the first RED to detect a recent PC BA, without access to the training set. We are the first to reverse-engineer BPs for PCs. We also proposed a novel detection statistic that conquers the intrinsic backdoor problem.

\vfill\pagebreak

\bibliographystyle{IEEEbib}
\bibliography{refs}

\vfill\pagebreak


\end{document}